\documentclass[%
 reprint,
 superscriptaddress,
 amsmath,amssymb,
 aps,
]{revtex4-1}

\usepackage{graphicx}
\usepackage{dcolumn}
\usepackage{bm}
\usepackage{xcolor}
\usepackage{url}

\usepackage{subfigure}
\usepackage{multirow}

\begin{document}


\title{Tackling the kaon structure function at EicC}

\author{Gang Xie}
\affiliation{Guangdong Provincial Key Laboratory of Nuclear Science, Institute of Quantum Matter, South China Normal University, Guangzhou 510006, China}
\affiliation{Institute of Modern Physics, Chinese Academy of Sciences, Lanzhou 730000, China}

\author{Chengdong Han}
\affiliation{Institute of Modern Physics, Chinese Academy of Sciences, Lanzhou 730000, China}
\affiliation{School of Nuclear Science and Technology, University of Chinese Academy of Sciences, Beijing 100049, China}

\author{Rong Wang}
\email{rwang@impcas.ac.cn}
\affiliation{Institute of Modern Physics, Chinese Academy of Sciences, Lanzhou 730000, China}
\affiliation{School of Nuclear Science and Technology, University of Chinese Academy of Sciences, Beijing 100049, China}

\author{Xurong Chen}
\email{xchen@impcas.ac.cn}
\affiliation{Guangdong Provincial Key Laboratory of Nuclear Science, Institute of Quantum Matter, South China Normal University, Guangzhou 510006, China}
\affiliation{Institute of Modern Physics, Chinese Academy of Sciences, Lanzhou 730000, China}
\affiliation{School of Nuclear Science and Technology, University of Chinese Academy of Sciences, Beijing 100049, China}


\date{\today}

\begin{abstract}
Measuring the kaon structure beyond the proton and pion structures is
one of the hot topics in hadron physics, as it is one way to understand the nature
of Nambu-Goldstone boson of QCD and to see the interplay
between the EHM mechanism and the HB mechanism for hadron mass generation.
In this paper, we present a simulation of the leading $\Lambda$ baryon tagged
deep inelastic scattering experiment at EicC (Electron-ion collider in China),
which is engaged to unveil the internal structure of kaon through Sullivan process.
According to our simulation results, the suggested experiment will cover the
kinematical domain of $0.05\lesssim x_{\rm K} \lesssim 0.85$ and $Q^2$ up to 50 GeV$^2$,
with the acceptable statistical uncertainties.
In the relatively low-$Q^2$ region ($<10$ GeV$^2$),
the Monte-Carlo simulation shows a good precision of the measurement ($<5$\%)
for the kaon structure function $F_2^{\rm K}$.
In the high-$Q^2$ region (up to 50 GeV$^2$), the statistical uncertainty of $F_2^{\rm K}$ is
also acceptable ($<10$\%) for the data at $x_{\rm K}<0.8$.
To perform such an experiment at an electron-ion collider,
a high-performance zero-degree calorimeter is required.
\end{abstract}

\maketitle

\section{Introduction}
\label{sec:intro}

The majority of the Universe's visible mass exist in the form of hadronic matter,
and the underlying theory of hadrons is quantum chromodynamics (QCD).
How the hadron acquires its mass is a fundamental and profound question
\cite{Ji:1994av,Ji:2021pys,Lorce:2017xzd,Roberts:2019ngp,Roberts:2020udq,Cui:2020dlm,Chen:2020ijn},
which is closely related to the confinement and the hadron structure.
The color confinement and the nonperturbative structure of the hadrons
are the peculiar challenging questions which attract a lot of interests.
Studying the meson structure provides a new and excellent direction to understand
the QCD predictions, since a meson is a simple object made of a quark
and an anti-quark in the quark model.

There are two mass generating mechanisms for the hadron:
Higgs Boson (HB) mechanism for the current quark mass
and Emergent Hadron Mass (EHM) mechanism for the complex interactions of quarks and gluons
\cite{Roberts:2021nhw,Arrington:2021biu}.
Dynamical chiral symmetry breaking is one of the features of QCD theory
\cite{Roberts:1994dr,Hawes:1993ef,Maris:1997tm}.
Based on the mass function of gluon from QCD's Schwinger function,
the gluon acquires a mass scale of $m_0\sim 0.43$ GeV, at zero momentum
\cite{Binosi:2016nme,Rodriguez-Quintero:2018wma,Cui:2019dwv}.
In the infrared region, the quark also becomes heavy by radiations and
absorptions of the gluons with the effective mass.
A vivid metaphor is that the quark dresses up with gluons
and turns into the constituent quark.
For the proton mass decomposition the chiral-limit mass
from dynamical chiral symmetry is dominant, while for the kaon mass decomposition
the interference between HB and EHM plays a dominant role with no chiral-limit mass.
Kaon is the Nambu-Goldstone boson mode of QCD,
thus the kaon is massless if the chiral symmetry is non-explicitly broken.
In the real world, the masses of the dressed quarks in kaon are
largely canceled by the attraction potential based on the wave function calculation
of two-body bound state \cite{Maris:1997hd,Roberts:2019ngp,Roberts:2020udq,Roberts:2021nhw}.

Along with the emergence of the kaon mass,
Dyson-Schwinger equations (DSE) predict a broadening quark
distribution function for the light quarks at the hadronic scale
(a quite low scale $Q_0^2$ where only the valence components of a hadron
are resolved by a probe) \cite{Chang:2013nia,Shi:2015esa,Raya:2019vwr,Cui:2020dlm,Ding:2019lwe,Cui:2020tdf}.
Compared to the up valence quark distribution, the strange valence quark
distribution is narrower, due to the heavier mass from HB mechanism.
Measuring the kaon structure will give a critical test on
this significant HB modulation of the EHM strange quark distribution \cite{Cui:2020dlm,Cui:2020tdf}.
To completely understand the EHM phenomenon, we should answer simultaneously
why the proton mass is heavy while the pion and kaon masses are light.
Investigating the kaon structure provides a clear way to see the interplay
between HB and EHM, due to the large coupling of strange quark to Higgs boson.
Moreover, the kaon structure measurement will test the fruitful calculations
from the nonperturbative approaches such as the continuum phenomenology of DSE
\cite{Shi:2015esa,Cui:2020dlm,Cui:2020tdf,Nguyen:2011jy,Chen:2016sno,Shi:2018mcb}
and lattice QCD (LQCD) \cite{Zhang:2017zfe,Chen:2019lcm,Lin:2020ssv,Zhang:2020gaj,Alexandrou:2020gxs}.

In experiment, the proton structure function has been measured precisely
with the help of high energy lepton beams or colliders worldwide.
However, the experimental data on the kaon structure function are extremely scarce.
There are only eight data points related to the quark distribution inside kaon,
which are accessed via the kaon-induced Drell-Yan process
of NA3 experiment at CERN more than fourty years ago \cite{Saclay-CERN-CollegedeFrance-EcolePoly-Orsay:1980fhh}.
Therefore, more and more experimental projects are proposed,
aiming for the better understandings on the kaon structure
and the EHM mechanism in the pseudoscalar meson sector.
On the AMBER facility at CERN, the implementation of kaon beam will provide
an extraction of the parton distribution functions (PDFs) of real kaon from
the Drell-Yan reaction \cite{Adams:2018pwt}. With the upgrade, the precision of the data at AMBER
will supreme that of NA3 data. At JLab of 12GeV upgrade,
the $\Lambda$ tagged deep inelastic scattering (DIS) process will be exploited
to study the structure of the virtual kaon \cite{TDIS,TDIS-kaon}.
This approach is similar to the leading neutron tagged DIS performed at HERA decades ago
\cite{Aaron:2010ab,Chekanov:2002pf},
for the determination of the pion structure function.
To acquire the kaon structure over a wide range of $Q^2$ and $x_{\rm K}$,
a high center-of-mass (c.m.) energy of the scattering is required.
Thus the electron-ion collider in US (US-EIC) \cite{AbdulKhalek:2021gbh,Accardi:2012qut,Aguilar:2019teb}
and in China (EicC) \cite{Chen:2018wyz,Chen:2020ijn,Anderle:2021wcy}
will provide the good opportunities for realization of this goal.

As there is almost no experimental data on kaon structure,
there is also no global analysis of the kaon PDFs.
Nevertheless in our previous work \cite{Han:2020vjp,IMParton-github},
we have determined the kaon PDFs from a model-dependent analysis of the eight data points
of NA3 experiment \cite{Saclay-CERN-CollegedeFrance-EcolePoly-Orsay:1980fhh},
based on the dynamical parton distribution model.
According to the JAM analysis of pion PDFs \cite{Barry:2018ort}, the addition of the leading neutron
tagged DIS data of H1 \cite{Aaron:2010ab} and ZEUS \cite{Chekanov:2002pf}
significantly reduce the uncertainties of sea quark and gluon distributions.
Similarly, the leading $\Lambda$ tagged DIS data in the future
will help fixing the sea quark and gluon distributions of the kaon.

Now there are ongoing discussions on building a polarized electron-ion collider in China,
by adding an electron beam to the high-intensity heavy ion accelerator facility \cite{Chen:2018wyz,Chen:2020ijn,Anderle:2021wcy}.
The optimal c.m. energy of the collision at EicC will be around 17 GeV \cite{Anderle:2021wcy}.
This would provide an excellent opportunity to probe the kaon structure
in the range of $0.02\lesssim x_{\rm K} \lesssim 1$,
i.e., from the sea quark region to the valence quark region.
Judged by the c.m. energy, EicC bridges well the measurement at JLab-12GeV \cite{TDIS,TDIS-kaon}
and the measurement at US-EIC \cite{AbdulKhalek:2021gbh,Accardi:2012qut,Aguilar:2019teb},
which will play an essential role in full mapping of the kaon structure.
The AMBER facility at CERN \cite{Adams:2018pwt} will run at the similar c.m. energy
of EicC, but the measurement is on the Drell-Yan reaction,
which is different from the tagged DIS process.
Hence the direct comparison between AMBER data and EicC data at the similar scale
will cross check each other and provide us a more definitive conclusion
of the kaon structure in the sea quark and valence quark region.
In this work, we suggest a kaon structure experiment at EicC
via the leading $\Lambda$ baryon tagged DIS process.
The feasibility and impact of the experiment will be demonstrated based on a simulation.
This work is similar to our previous simulation study of the pion structure at EicC \cite{Xie:2020uck}.

The organization of the paper is as follows.
The under-discussion EicC is briefly introduced in Sec. \ref{sec:EicC}.
The model for the leading $\Lambda$ baryon tagged DIS
is described in Sec. \ref{sec:LambdaTaggedDIS}.
The input kaon PDFs for the simulation is illustrated in Sec. \ref{sec:kaonPDFs}.
The invariant kinematic and final-state kinematic distributions
of the Monte-Carlo simulation are shown in Sec. \ref{sec:KineDistributions}.
The error projections of the proposed kaon structure function measurement are
shown in Sec. \ref{sec:F2kaonErrorProjection}.
Finally, some discussions and a concise summary is given in Sec. \ref{sec:summary}.

\section{Electron-ion collider in China}
\label{sec:EicC}

The proposed polarized electron-ion collider in China is
a future high energy nuclear physics project,
aiming at the precise measurement of the nucleon structure in the sea quark region,
the exotic hadron physics, the nuclear matter effect, etc.
EicC will cover the variable c.m. energies from 15 to 20 GeV,
with the luminosity above $10^{33}$ cm$^{-2}$s$^{-1}$ \cite{Anderle:2021wcy}.
In this work, we assume EicC runs with the electron beam energy of 3.5 GeV
and the proton beam energy of 20 GeV.
The luminosity of EicC is around 100 times of the previous HERA collider in Germany \cite{Aaron:2010ab,Chekanov:2002pf}.
Therefore EicC will provide much precision data on the sea quark structures of the hadrons.
EIC in US will focus on the gluon dominant region.
In the future, much more details will be unveiled by the new facilities in the high precision era.

For the conceptual design of EicC, the central detector and end-cap detector systems will be
constructed inside and around the solenoid magnet with the cutting-edge technologies \cite{Anderle:2021wcy}.
The far-forward detector complex of high performance will also be implemented
in both beam directions, such as the Roman pot inside the beam pipe,
the off-momentum detectors around the beam line,
and the zero-degree calorimeter (ZDC).
Thus the EicC facility will provide us a good opportunity to tag the high energy $\Lambda$ baryon
of high pseudorapidity, so as to measure the leading $\Lambda$ tagged DIS.
To reconstruct the $\Lambda$ from the decay proton and $\pi^{-}$ encounters a lot of difficulties,
due to the deflections of the charged particles by the complicated magnets around the beam pipe
and the challenge of particle identifications of high energy electron, pion, kaon and proton.
It is more doable to reconstruct the $\Lambda$ baryon by measuring its neutral decay
(neutron and $\pi^0$) with ZDC.
Identifications of neutron and $\pi^0$ could be realized by differentiating
the hadronic shower and the electromagnetic shower.

\section{Leading $\Lambda$ tagged DIS and kaon structure function}
\label{sec:LambdaTaggedDIS}

\begin{figure}[htbp]
\centering
\includegraphics[scale=0.34]{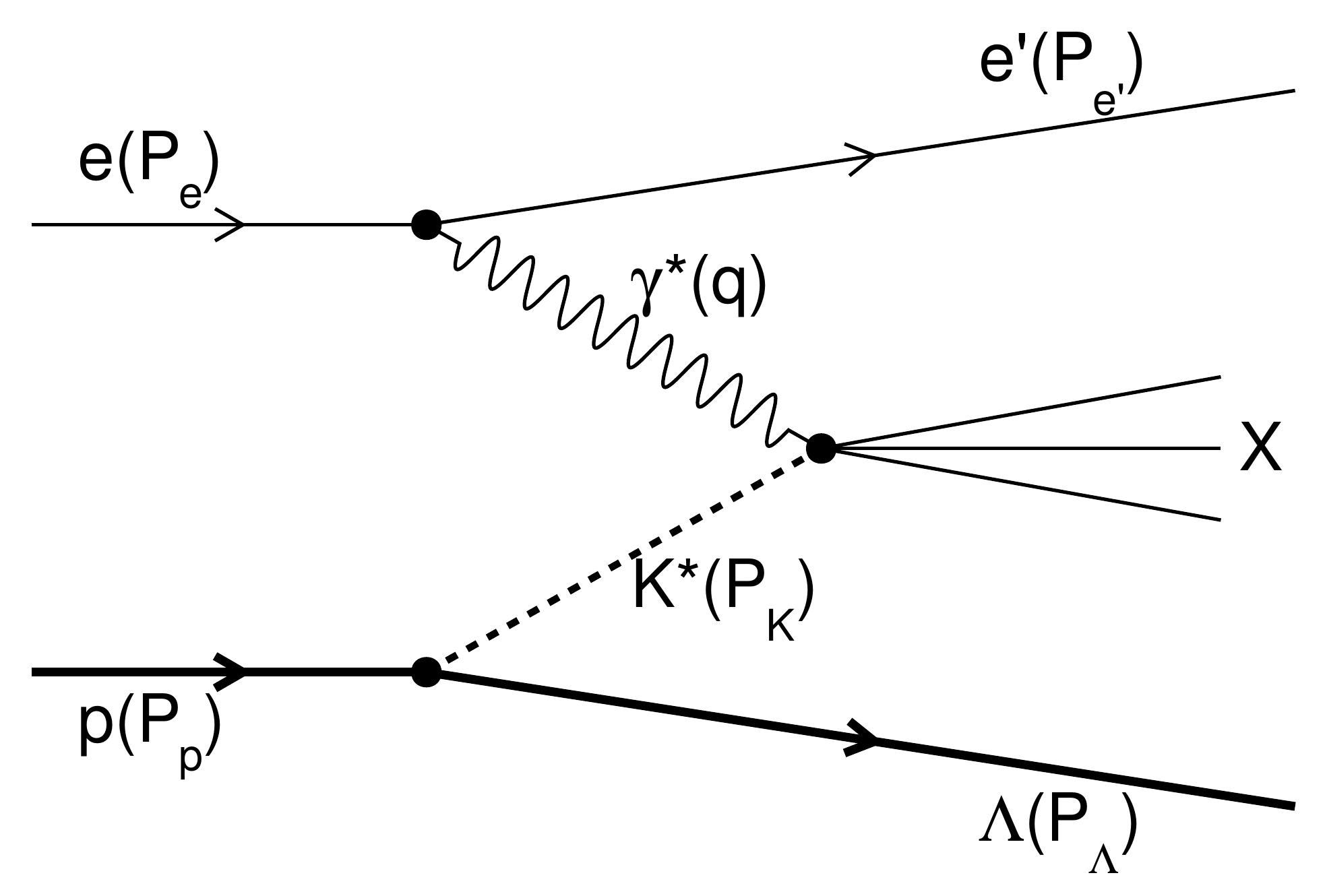}
\caption{The Sullivan process \cite{Sullivan:1971kd} for deep inelastic scattering with
the production of a leading $\Lambda$ baryon.
The leading $\Lambda$ carries a great amount of the momentum of the beam proton.  }
\label{fig:LLambdaDIS-diagram}
\end{figure}

To probe the kaon structure in high-energy $e-p$ collision,
we exploit the abundant ``kaon cloud'' from proton dissociation
due to the large coupling $g_{N\Lambda K}$.
This type of electron-``meson cloud'' scattering dominates in the $t$ channel
with one meson exchange, called the Sullivan process \cite{Sullivan:1971kd},
which is shown in Fig. \ref{fig:LLambdaDIS-diagram}.
To make sure the electron beam hitting the ``kaon cloud'',
we need to tag the leading $\Lambda$ of high energy and small transverse momentum.
The $\Lambda$ baryon acts as the spectator carrying a large fraction of the incoming
proton's momentum and going far-forward.
To measure the kaon structure, we need also to make sure the virtual kaon is
broken up by the high energy probe.
In the literature, the internal structure of the quasi-real kaon in the process
resembles the internal structure of the real kaon,
as long as the momentum transfer is not large ($\lesssim 0.9$ GeV$^2$) \cite{Qin:2017lcd}.

The invariant kinematical variables describing the leading $\Lambda$ tagged DIS are:
the momentum square of the photon probe $Q^2$, the Bjorken variable $x_B$,
the inelasticity $y$ of the scattering, the longitudinal momentum fraction $x_{\rm L}$
carried by the $\Lambda$ baryon, and the square of the momentum transfer
from the proton to the virtual kaon $t$.
According to the momenta of the particles labeled in Fig. \ref{fig:LLambdaDIS-diagram},
these kinematical variables are defined as,
\begin{equation}
\begin{split}
Q^2 \equiv -q^2,\\
x_{\rm B} \equiv \frac{Q^{2}}{2P_{\rm p} \cdot q},\\
y \equiv \frac{P_{\rm p} \cdot q}{P_{\rm p} \cdot P_{\rm e}},\\
x_{\rm L} \equiv \frac{P_{\rm \Lambda}\cdot q}{P_{\rm p}\cdot q},\\
t \equiv (P_{\rm p} - P_{\rm \Lambda})^2 = p_{\rm K}^2.
\end{split}
\label{eq:DIS-kine-definition}
\end{equation}
In definition, $x_{\rm L}$ denotes the longitudinal momentum fraction (energy fraction approximately)
of the final $\Lambda$ baryon to the incoming proton.
In the DIS experiment, the leading $\Lambda$ tagged process dominates
in the large-$x_L$ region ($\gtrsim 0.5$),
hence a proper cut on the $x_L$ variable selects efficiently
the events that are sensitive to the kaon structure.
$t$ denotes momentum square of the virtual kaon,
which is an important variable for the extrapolation of the real kaon structure.

In order to estimate the statistical error of the measurement,
we need to know the number of events of interest.
Therefore we need first to know the cross section of the reaction.
With the azimuthal angle integrated, the four-fold differential cross section
of the leading $\Lambda$ tagged DIS is written as \cite{Holtmann:1996ac,Chekanov:2002pf,Aaron:2010ab},
\begin{equation}
\begin{split}
&\frac{d^4\sigma({\rm ep\rightarrow e\Lambda X})}{dx_{\rm B}dQ^2dx_{\rm L}dt} =
\frac{4\pi\alpha^2}{x_{\rm B}Q^4}\left(1-y+\frac{y^2}{2}\right)F_2^{\rm L\Lambda(4)}(Q^2, x_{\rm B}, x_{\rm L}, t)\\
&= \frac{4\pi\alpha^2}{x_{\rm B}Q^4}\left(1-y+\frac{y^2}{2}\right) F_2^{\rm K}\left(\frac{x_{\rm B}}{1-x_{\rm L}} ,Q^2\right)f_{\rm K^+/p}(x_{\rm L},t).
\end{split}
\label{eq:DiffXSection-four}
\end{equation}
From this equation, we see that we can extract the four fold
leading-$\Lambda$ structure function $F^{\rm L\Lambda(4)}_2$.
In the kaon pole model, the leading-$\Lambda$ structure function can be factorized
into the product of the kaon structure function $F_2^{\rm K}$
and the kaon flux around the proton $f_{\rm K^+/p}$.
In a effective theory of kaon pole,
the kaon flux is given by \cite{Holtmann:1996ac,Chekanov:2002pf,Aaron:2010ab},
\begin{equation}
\begin{split}
& f_{\rm K^+/p}(x_{\rm L},t)=\\
& \frac{1}{2\pi}\frac{g^2_{\rm N\Lambda K}}{4\pi}(1-x_{\rm L})\frac{-t}{(m_{\rm K}^2-t)^2}
{\rm exp}\left(-R^2_{\rm \Lambda K}\frac{t-m_{\rm K}^2}{1-x_{\rm L}}\right),
\end{split}
\label{eq:pion-flux}
\end{equation}
in which the coupling is $g_{\rm N\Lambda K}^2/4\pi = 14.7$,
and $R_{\rm \Lambda K} = 1$ GeV$^{-1}$ is a form-factor parameter
represents the radius of the $\Lambda-K$ Fock state of the proton.
With these formulae above, we can compute the cross section of
the leading $\Lambda$ baryon tagged DIS process.

\section{Parton distribution functions of kaon}
\label{sec:kaonPDFs}

\begin{figure}[htbp]
\centering
\includegraphics[scale=0.42]{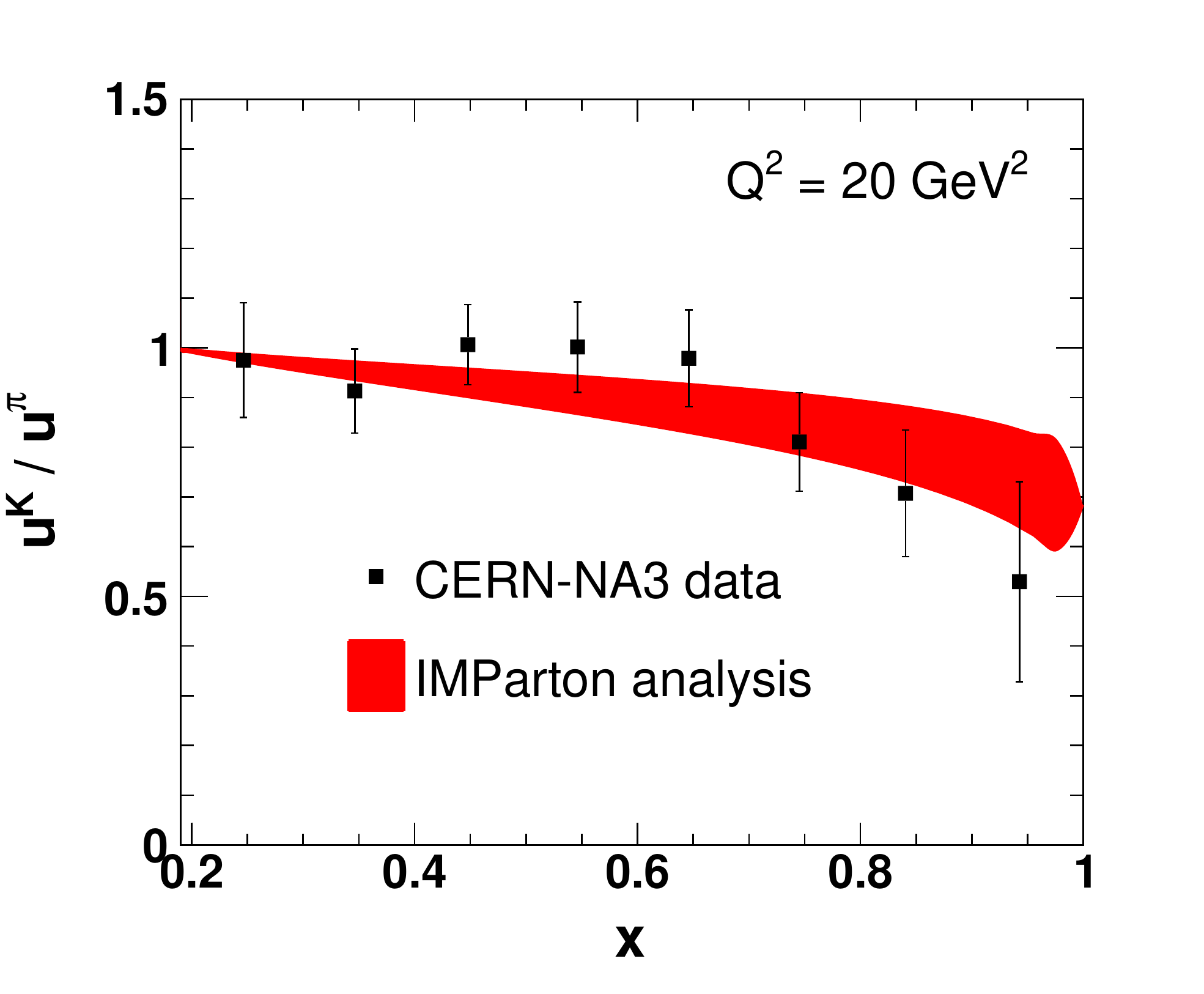}
\caption{Comparison of the model predicted ratio $u^{\rm K}/u^{\rm \pi}$ \cite{Han:2020vjp}
as a function of $x$ with CERN-NA3 experimental data \cite{Saclay-CERN-CollegedeFrance-EcolePoly-Orsay:1980fhh}.  }
\label{fig:ratio-kaon-pion}
\end{figure}

For estimating the cross section and building a event generator for the process,
the last input is the PDFs of kaon.
Since there is almost no experimental data on kaon structure,
there is no global analysis on the kaon PDFs.
Therefore in this work, we use the kaon PDFs provided by an model-dependent
analysis of the NA3 data only \cite{Han:2020vjp}.
This analysis is based on the dynamical parton distribution model
with only valence quark distributions at $Q_0^2$, where the sea quark
and gluon distributions are completely produced from the QCD fluctuations.
Fig. \ref{fig:ratio-kaon-pion} shows the ratios of the kaon $\bar{u}$ distribution
to the pion $\bar{u}$ distribution, compared to the NA3 data.
The experimental data indicate that the kaon up quark distribution
is lower that the pion up quark distribution in the valence region.
We see that the used kaon PDFs in this work are consistent with
the only experimental data obtained decades ago \cite{Saclay-CERN-CollegedeFrance-EcolePoly-Orsay:1980fhh}.

\begin{figure}[htbp]
\centering
\includegraphics[scale=0.4]{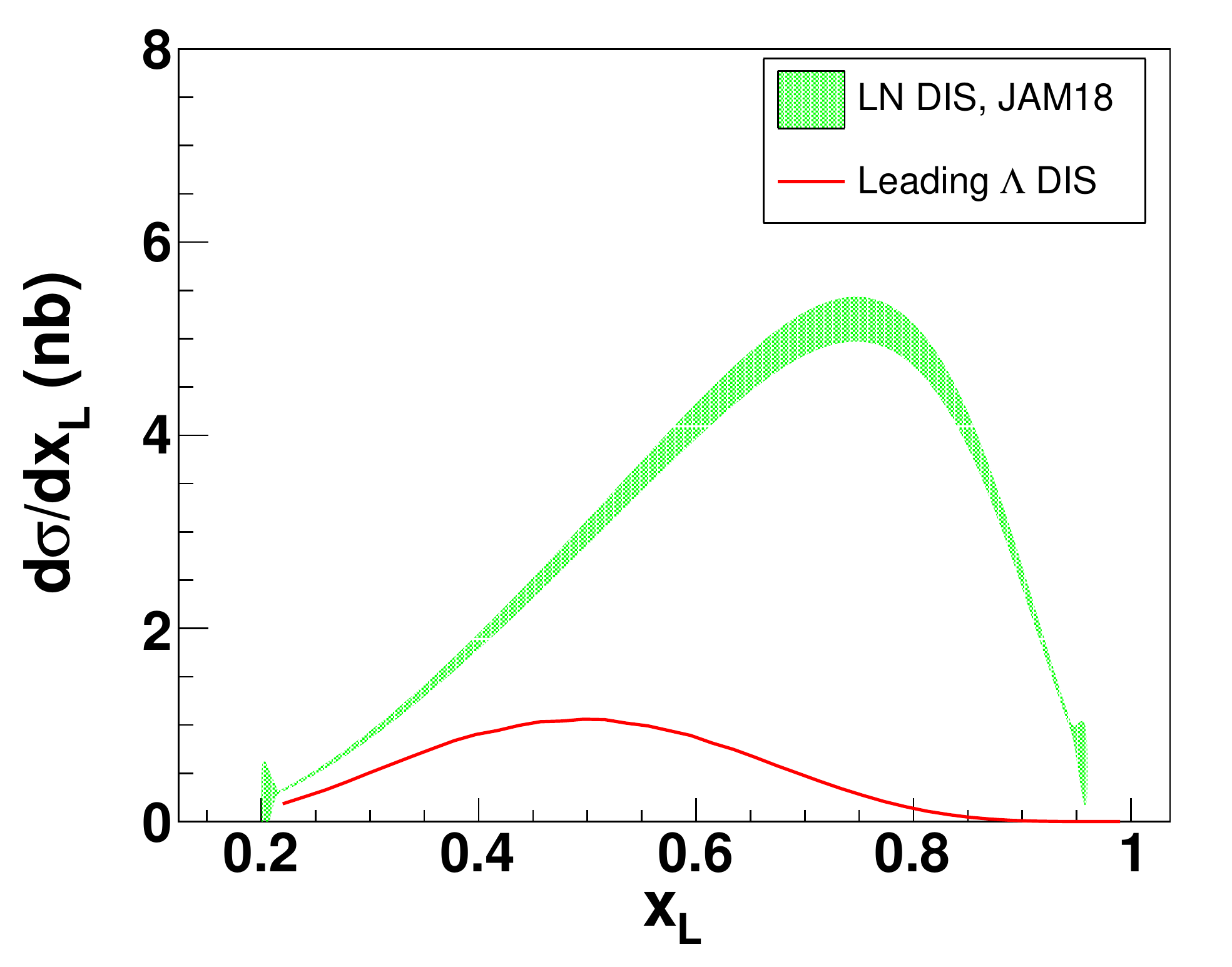}
\caption{The one-fold differential cross section as a function of the
longitudinal momentum fraction $x_{\rm L}$ of the far-forward $\Lambda$.
The cross section of leading $\Lambda$ tagged DIS is based on the model described in this work.
The cross section of leading neutron tagged DIS is taken from the reference \cite{Xie:2020uck} for comparison,
which is based on the JAM pion PDFs \cite{Barry:2018ort}.   }
\label{fig:dsigma_dxl}
\end{figure}

With the cross section model described in last section
and the kaon PDFs, we calculate the differential cross
section of leading $\Lambda$ tagged DIS as a function of $x_{\rm L}$,
which is displayed in Fig. \ref{fig:dsigma_dxl}.
The experimental data and the model prediction for the differential cross
section of leading neutron tagged DIS process are also shown in the figure.
We see that the cross section of leading $\Lambda$ tagged process
is much smaller than that of leading neutron tagged process.
The other finding is that the cross section of leading $\Lambda$ tagged
DIS dominates in relatively lower $x_{\rm L}$ region.
This is because the ``kaon cloud'' (of heavier mass than the ``pion cloud'')
also carries a decent amount of the momentum of the beam proton.
Hence in the analysis, we should a lower cut on $x_{\rm L}$
to select the events of interests.

\section{Distributions of invariant and final-state kinematics}
\label{sec:KineDistributions}

Following the theoretical framework discussed in the above two sections,
we develop a event generator program of leading $\Lambda$ tagged DIS process.
In the simulation, the electron beam energy is taken to be 3.5 GeV
and the proton beam energy is taken to be 20 GeV.
The $z$ direction of coordinate is chosen to be the momentum of the incoming proton beam.
In order to efficiently generate the events in the kinematic region of interests,
we set the following ranges of kinematics in the Monte-Carlo simulation:
$x_{\rm B,min}<x_{\rm B}<1$, $1~{\rm GeV}^2 < Q^2 < 50~{\rm GeV}^2$,
$0.01~{\rm GeV}^2 < -t < 1~{\rm GeV}^2$, and $0.5 < x_{\rm L} < 1$.

\begin{figure}[htbp]
\centering
\includegraphics[scale=0.45]{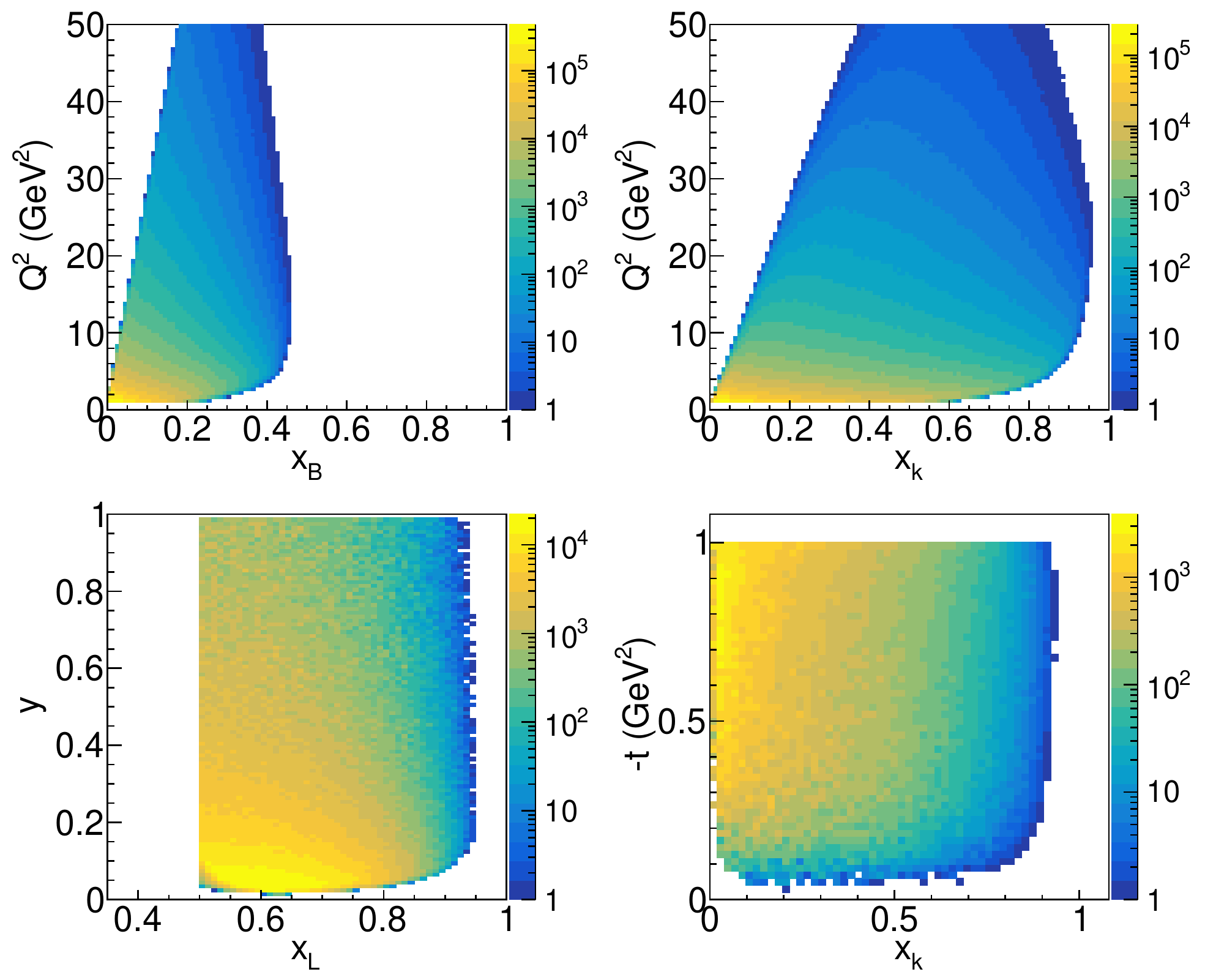}
\caption{The distributions of the invariant kinematical variables $Q^2$, $x_{\rm B}$, $x_{\rm K}$,
$x_{\rm L}$, $y$, and $-t$, for the simulation data of $\Lambda$ baryon tagged DIS at EicC.  }
\label{fig:invar_kine}
\end{figure}

\begin{figure}[htbp]
\centering
\includegraphics[scale=0.45]{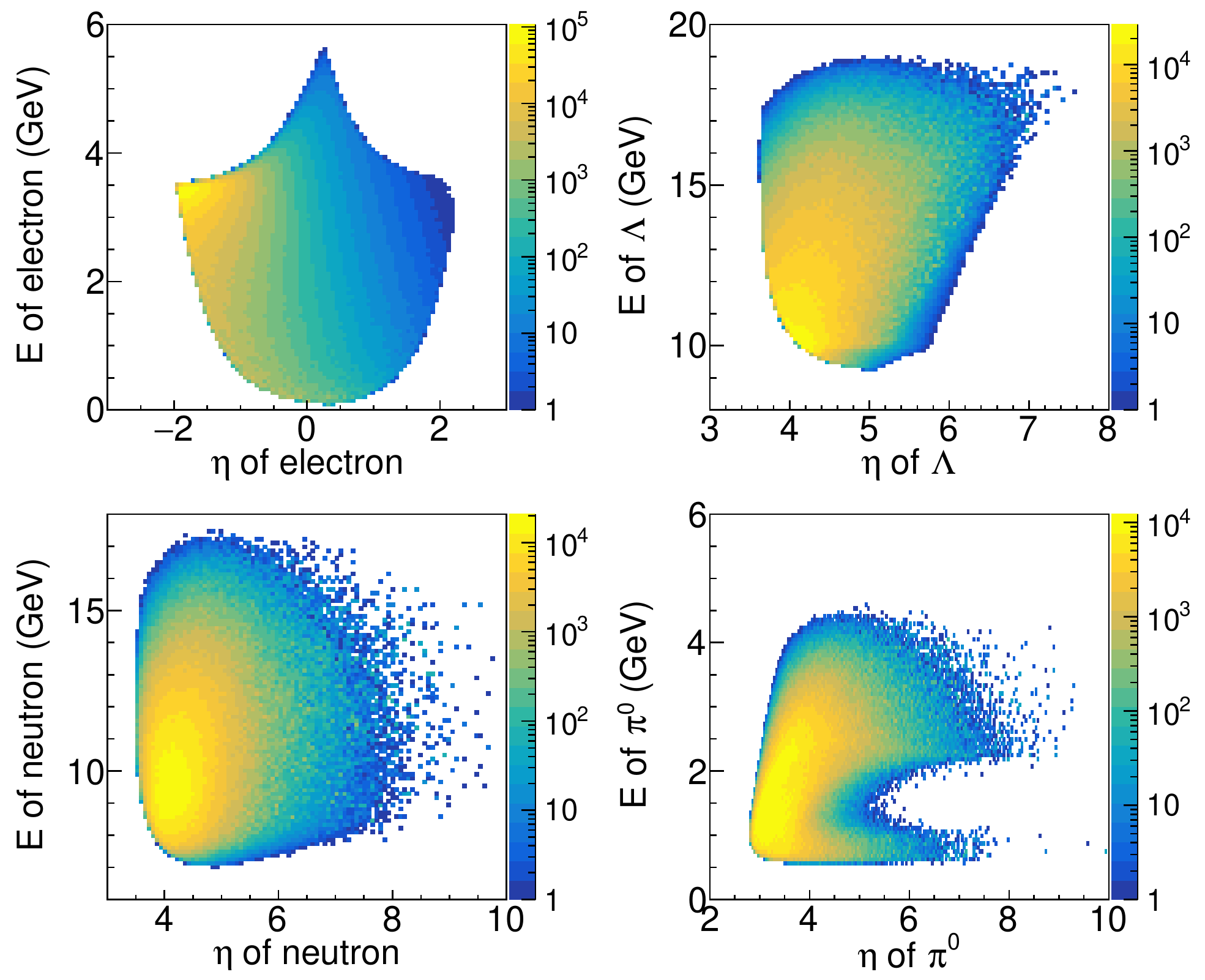}
\caption{The Monte-Carlo simulated energy and pseudorapidity distributions of the measured final-state particles:
electron and $\Lambda$. The distributions of the decays (neutron and $\pi^0$) of $\Lambda$ are also shown.   }
\label{fig:final_kine_E_eta}
\end{figure}

Fig. \ref{fig:invar_kine} shows the cross-section weighted invariant kinematic
distributions of the leading $\Lambda$ tagged DIS events simulated,
which are projected in two dimensional spaces.
The events are mainly distributed in low $Q^2$,
small $x_{\rm B}$, small $x_{\rm K}$, and small $y$ region.
Fig. \ref{fig:final_kine_E_eta} shows the energy and pseudorapidity distributions
of the final-states: electron, $\Lambda$, and the neutral
decays of $\Lambda$ (neutron and $\pi^0$).
We see that all the scattered electrons can be collected with the central detectors at EicC,
while the leading $\Lambda$ of high energy and large rapidity can only be 
detected with the forward detectors.
The decay neutron of $\Lambda$ mainly distribute around the pseudorapidity of 5.
The decay $\pi^0$ of $\Lambda$ mainly distribute around the pseudorapidity of 3.5.
There is a small portion of $\pi^0$ going to the end-cap detector system.

\begin{figure}[htbp]
\centering
\includegraphics[scale=0.45]{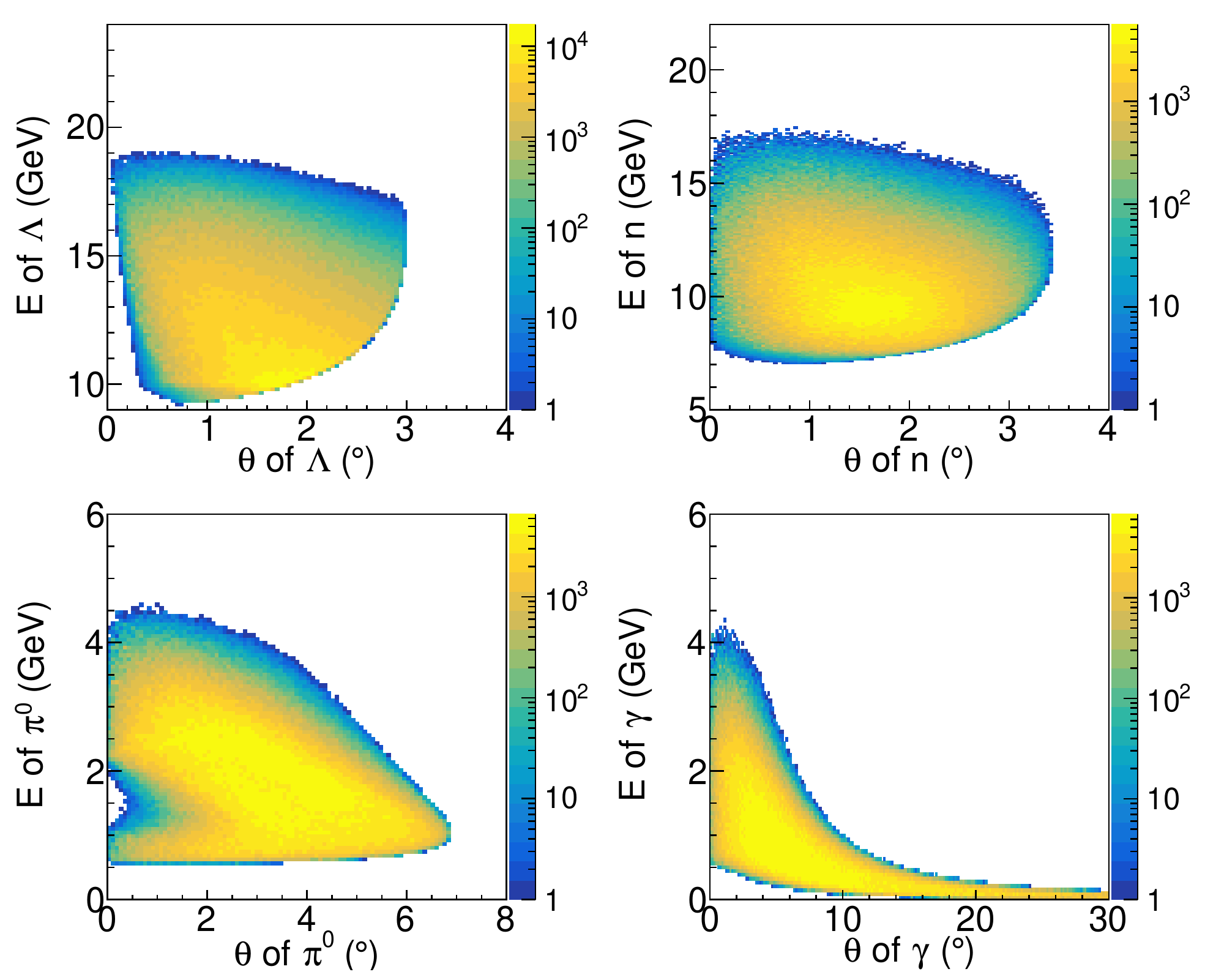}
\caption{The energy and $\theta$ angle distributions of high momentum far-forward $\Lambda$ baryon
and its decay chains, for the simulation data of $\Lambda$ baryon tagged DIS at EicC.    }
\label{fig:final_kine_E_theta}
\end{figure}

To investigate with more details on the forward $\Lambda$ decay,
we show the distributions of the two photons from $\pi^0$ decay as well.
Fig. \ref{fig:final_kine_E_theta} shows the energy and $\theta$ angle distributions
of $\Lambda$, $n$, $\pi^0$, and $\gamma$.
We see that most of the neutrons from $\Lambda$ decay go to ZDC.
However, some photons of low energy from $\pi^0$ decay go to the central
and end-cap detectors, and some photons of high energy $\pi^0$ decay go to ZDC.
To get rid of the background noise of the detector,
we set the low-energy threshold to 200 MeV for the high energy photon detection.
According to the conceptual design of EicC,
we choose the $\theta$ angle cut to be $\theta_{\gamma}<3^\circ$ or $6^\circ <\theta_{\gamma}<174^\circ$.

\begin{figure}[htbp]
\centering
\includegraphics[scale=0.38]{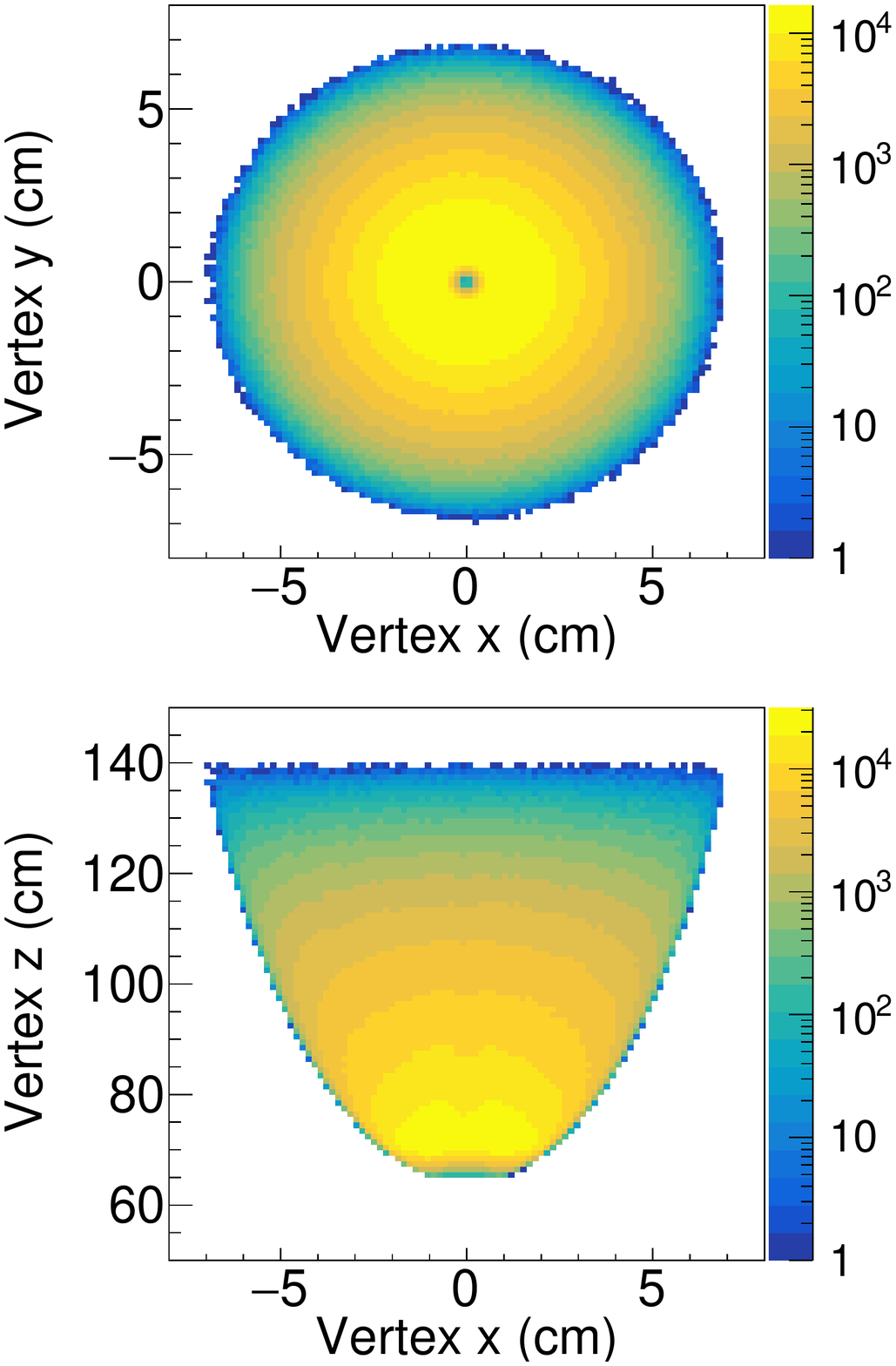}
\caption{The distributions of the decay vertex of the forward high-energy $\Lambda$ baryon,
for the simulation data of $\Lambda$ baryon tagged DIS at EicC.    }
\label{fig:Lambda_decay_vertex}
\end{figure}

The decay vertex of the leading $\Lambda$ from Sullivan process
is displayed in Fig. \ref{fig:Lambda_decay_vertex}, which could provide some guidance
for the future analysis of the vertex reconstruction.
The decay vertex is close to the beam line with a small transverse distance.
The decay vertex is about 80 cm from the production vertex,
which could be used as a cut to select the leading $\Lambda$ tagged DIS events.

\section{Statistical error projections of kaon structure function at EicC}
\label{sec:F2kaonErrorProjection}

\begin{figure}[htbp]
\centering
\includegraphics[scale=0.45]{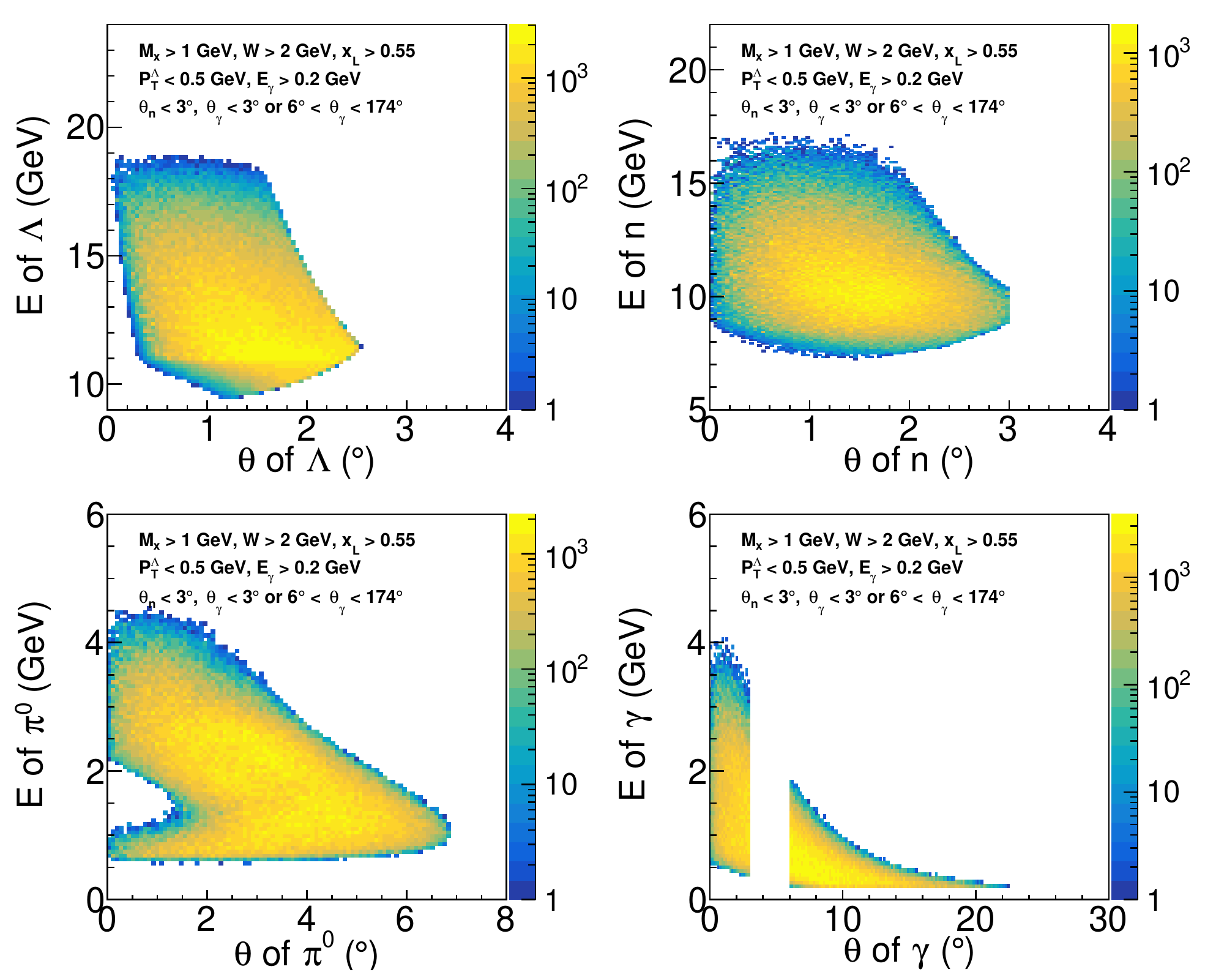}
\caption{The energy and $\theta$ angle distributions of high momentum far-forward $\Lambda$ baryon
and its decay chains, with the geometric cut and energy threshold of electromagnetic calorimeters applied,
for the simulation data of $\Lambda$ baryon tagged DIS at EicC.   }
\label{fig:final_kine_E_theta_after_cut}
\end{figure}

To estimate the statistical error of the kaon structure function,
we simply need to estimate the statistical error of the cross section,
since these two experimental observable are directly related.
The statistical uncertainty of the cross section measurement
depends on the number of events collected in the experiment.
To estimate the number of events of a experiment,
we need to know the cross section of the reaction
(provided by the model described above),
the integrated luminosity of the experiment,
and the event selection criteria of the reaction.
For a year running of good quality beams, EicC could accumulate about
50 fb$^{-1}$ integrated luminosity of $e-p$ collisions.
Hence we take the integrated luminosity of 50 fb$^{-1}$ for the simulation.
To make sure the collected events are mainly from electron-``kaon cloud'' collisions,
we take the following event selection criteria:
$x_{\rm L}>0.55$, $P_{\rm T}^{\rm \Lambda}<0.5$ GeV,
$M_{\rm X}>1$ GeV, $W>2$ GeV.
$x_{\rm L}>0.55$ and $P_{\rm T}^{\rm \Lambda}<0.5$ GeV ensure the events are from
Sullivan process of $t$ channel, and $W > 2$ GeV is the usual DIS criterium.
Fig. \ref{fig:final_kine_E_theta_after_cut} shows the energy and pseudorapidity
distributions of the $\Lambda$ and its decays, after the event selection criteria,
the geometrical acceptance of the detectors and the low energy threshold of the calorimeters.
The zero-degree calorimeter is required to cover the angle from 0 to 3 degrees around the beam,
to collect more neutrons and photons.

\begin{figure}[htbp]
\centering
\includegraphics[scale=0.4]{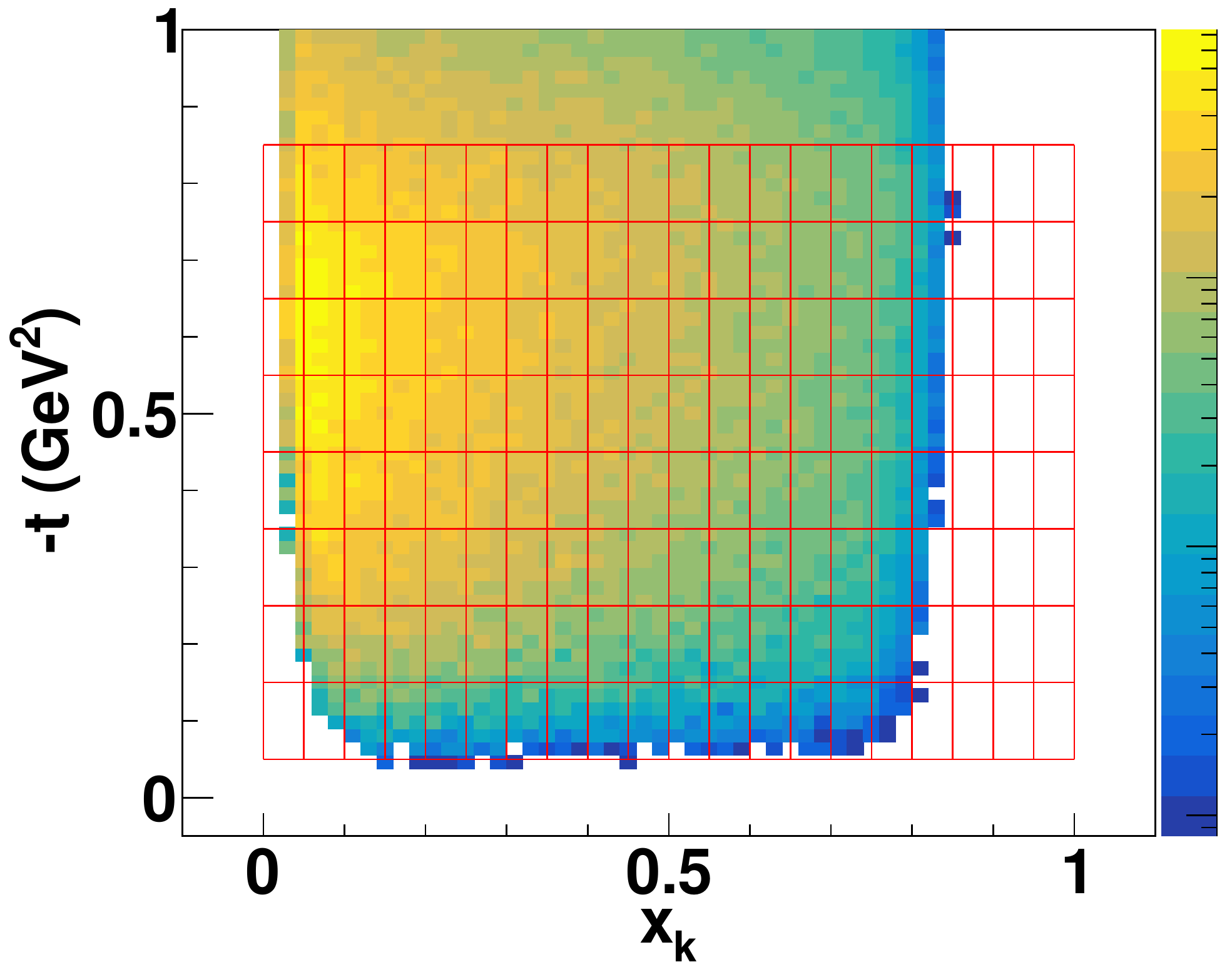}
\caption{ The binning scheme in $-t$ versus $x_{\rm K}$ plane for
the Monte-Carlo data in $Q^2$ range of $(3,5)$ GeV$^2$.  }
\label{fig:binning_Q2_3_to_5GeV2}
\end{figure}

To get the cross section at each kinematical point,
we need to count the number of events in different kinematical bins.
The typical kinematical binning is shown in Fig. \ref{fig:binning_Q2_3_to_5GeV2},
for the events in the $Q^2$ range of $(3,5)$ GeV$^2$.
We focus on the events at relatively small $|t|$ ($<0.85$ GeV$^2$),
a condition suggested by DSE calculation to make sure the extrapolation
to the real kaon structure valid and effective \cite{Qin:2017lcd}.
With the event selection criteria discussed in the above paragraph,
we calculate the number of events in each bin, with the following formula,
\begin{equation}
N_{\rm i} = L \overline{\sigma}_{\rm i} B_r \epsilon \Delta x_{\rm K} \Delta Q^2 \Delta x_{\rm L} \Delta t (1-x_{\rm L}),
\label{eq:number of events}
\end{equation}
where $L$ is the integrated luminosity of the suggested experiment,
$\overline{\sigma}_{\rm i}$ is the averaged differential cross section in kinematical bin $i$,
$B_r$ is the branching ratio of $\Lambda$ decaying into neutron and two photons,
$\epsilon$ is the detector efficiency for collecting all the final states of the reaction,
i.e. $\epsilon = \epsilon_n * \epsilon_{\gamma_1} * \epsilon_{\gamma_2}$,
the factor $(1-x_L)$ is the Jacobian coefficient for the transform from $x_{\rm B}$ space to $x_{\rm K}$ space,
and the other factors together express the size of the kinematical bins.
For the detectors of common performance, we assume $\epsilon_{\gamma} = 90 \%$
for detecting and identifying the photons from $\pi^0$ decay,
and $\epsilon_n = 50 \%$ for detecting and identifying the far-forward neutrons.
Finally the relative statistical error of kaon structure function $\delta(F_2^{\rm K}) / F_2^{\rm K}$
in each kinematical bin is estimated to be $1/\sqrt{N_{\rm i}}$.

\begin{figure}[htbp]
\centering
\includegraphics[scale=0.42]{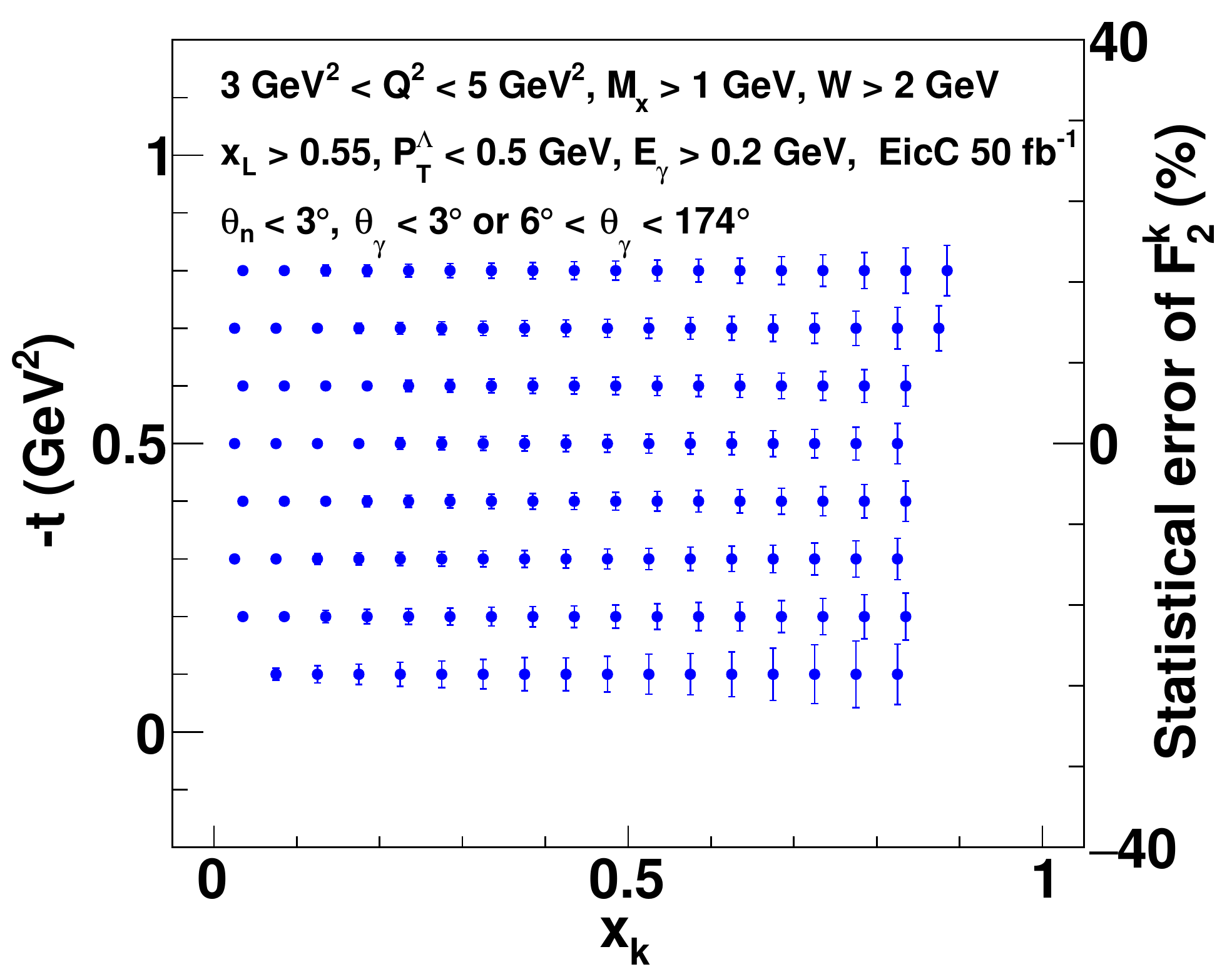}
\caption{The statistical error projections of the
kaon structure function at $Q^2 \sim 4$ GeV$^2$.
We calculate the statistical error at each bin center.
The right axis is a scale indicating how large the statistical error is.    }
\label{fig:F2k-error-Q2-3-5}
\end{figure}

\begin{figure}[htbp]
\centering
\includegraphics[scale=0.42]{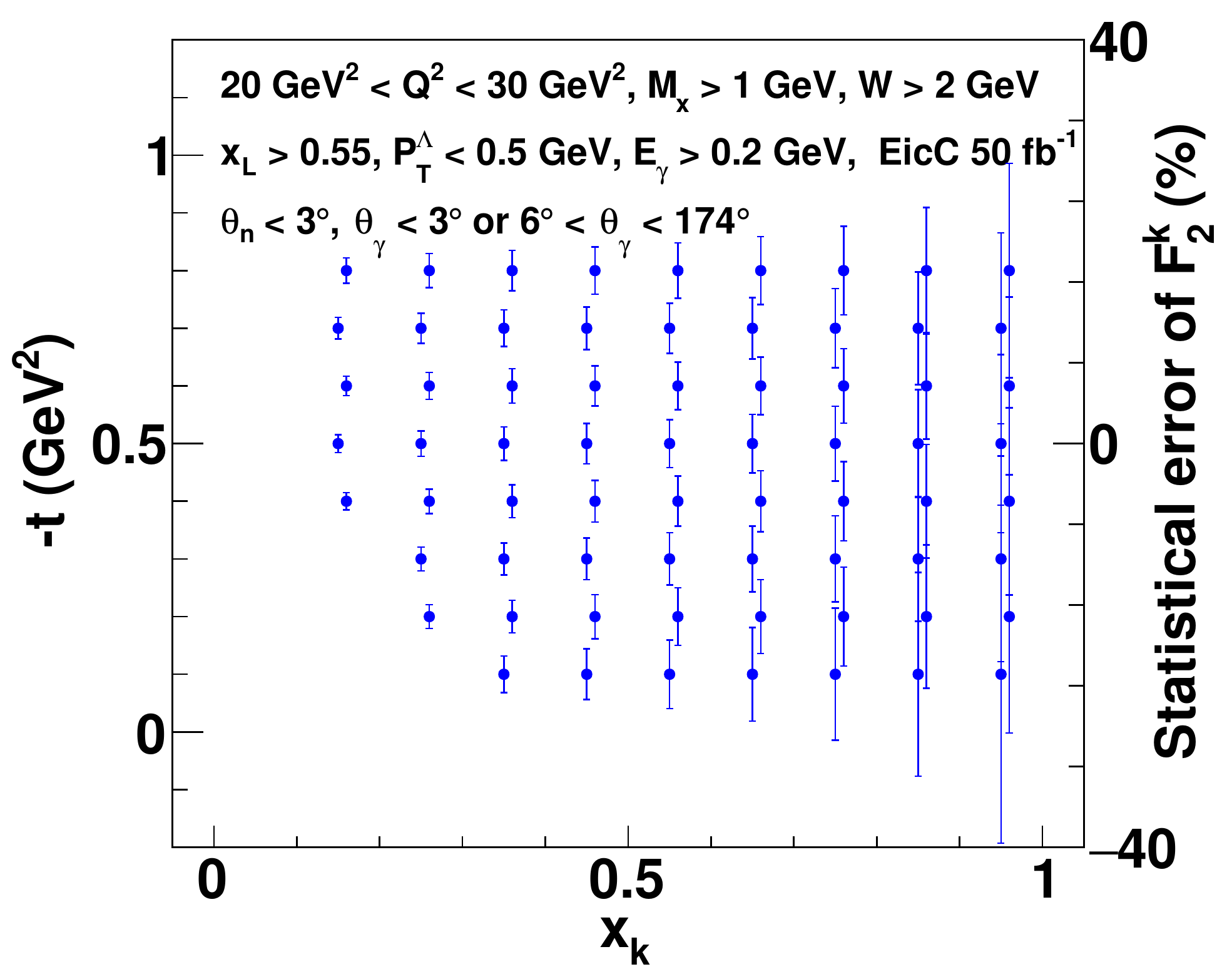}
\caption{The statistical error projections of the
kaon structure function at $Q^2 \sim 25$ GeV$^2$.
We calculate the statistical error at each bin center.
The right axis is a scale indicating how large the statistical error is.   }
\label{fig:F2k-error-Q2-20-30}
\end{figure}

\begin{figure}[htbp]
\centering
\includegraphics[scale=0.42]{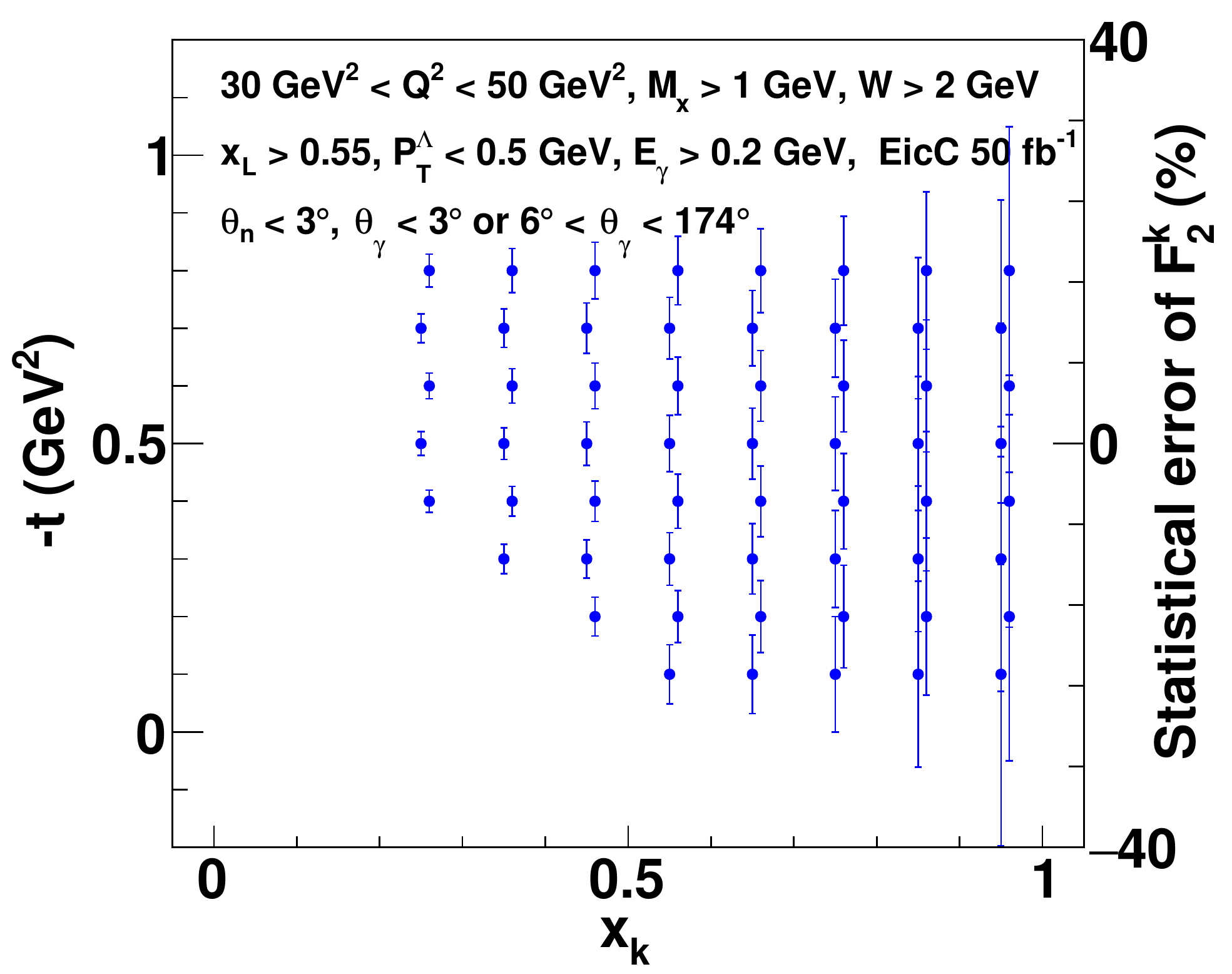}
\caption{The statistical error projections of the
kaon structure function at $Q^2 \sim 40$ GeV$^2$.
We calculate the statistical error at each bin center.
The right axis is a scale indicating how large the statistical error is.   }
\label{fig:F2k-error-Q2-30-50}
\end{figure}

By counting the simulated events in each kinematical bin,
we calculate the statistical uncertainty of the kaon structure function
for the proposed experiment at EicC.
Fig. \ref{fig:F2k-error-Q2-3-5} shows the relative statistical error of $F_2^{\rm K}$
in the kinematical bin of $3~{\rm GeV^2}<Q^2<5~{\rm GeV^2}$.
We see in the plot that the statistical uncertainty goes up with
the $x_{\rm K}$ increasing up.
For the data at $x_{\rm K}<0.3$, the projected statistical uncertainty is smaller than 1\%.
With the $x_{\rm K}$ increasing up around 0.85, the statistical uncertainty is around 5\%.
These precise data in the future will provide an excellent test
of the predictions of lattice QCD and DSE.
At higher $Q^2$ up to 50 GeV$^2$, the statistical uncertainty projections are also
projected and depicted in Fig. \ref{fig:F2k-error-Q2-20-30} ($Q^2\sim 25$ GeV$^2$)
and Fig. \ref{fig:F2k-error-Q2-30-50} ($Q^2\sim 40$ GeV$^2$).
With wider kinematical bins and fewer data points,
the estimated statistical precision of $F_2^{\rm K}$ measurement is still good.
For the data points in the region of $x_{\rm K}<0.6$,
the relative statistical uncertainties are less than 5\%.
And for the data points in the region of $x_{\rm K}<0.8$,
the relative statistical uncertainties are less than 10\%.
These experimental data over a wide range of $Q^2$ will provide us an interesting
opportunity to test the QCD evolution equations in the kaon sector,
and to extract the gluon distribution in the kaon via the scaling violation.

\section{Discussions and summary}
\label{sec:summary}

The leading $\Lambda$ tagged DIS experiment at EicC has been simulated
to study the kaon structure function.
The charged decays ($p$ and $\pi^-$) from $\Lambda$ of high energy are deflected by
the beam magnets and they are difficult to be separated from the beam protons.
Moreover, particle identifications of the high-energy charged particles
around the beam pipe and beam magnets are very challenging.
It is more doable to measure the neutral decays ($n$ and $\pi^0$) of $\Lambda$
with ZDC, since the neutral particles are not deflected by the beam magnets.
From our simulation, the high energy neutrons are collected with ZDC,
while the photons from $\pi^0$ decay are collected with both ZDC and end-cap electromagnetic
calorimeter in the interaction region of EicC.

We suggest the ZDC at EicC cover the $\theta$ angle from 0 to 3 degrees
in order to collect the forward neutrons as many as possible.
The angular resolution (position resolution) and energy resolution of ZDC
should be studied with more details in the future.
It is critical to has a small angular resolution for two reasons.
First, the small angular resolution of ZDC is important for
measuring the $t$-dependence of the cross section,
so as to perform a good extrapolation to the real kaon structure.
Second, we need a small angular resolution
to separate the clusters of the energy depositions of the multiple neutral particles.
For the shashlik calorimeter, the position resolution can be smaller than 1 cm,
which is quite promising.
The particle identification ability of ZDC is also important
for the success of the experiment.
The longitudinal profiles of the electromagnetic and hadronic showers are different.
It is a mature technology to differentiate the photons from the neutrons
with the implementation of a pre-shower.

From our simulation, the conclusion is that the kaon structure experiment at EicC is
feasible with the high-performance zero-degree calorimeter.
We have made the projections on the statistical errors of the kaon structure function
based on a cross section model of $\Lambda$ tagged DIS,
with an assumed integrated luminosity of 50 fb$^{-1}$
and the acceptances of the conceptual EicC detectors.
At the collision c.m. energy around 17 GeV,
EicC covers a broad kinematical range of $0.05\lesssim x_{\rm K} \lesssim 0.9$
and with the resolution scale $Q^2$ up to 50 GeV$^2$.
In the small $x_{\rm K}$ and low $Q^2$ region ($<10$ GeV$^2$),
the statistical uncertainty is smaller than 1\%.
At high $x_{\rm K}\sim 0.85$ and a low $Q^2\sim 4$ GeV$^2$,
the statistical uncertainty is just around 5\%.
At high $Q^2$ and with fewer kinematical bins,
the statistical uncertainties are less than 5\%
for the data points in the region of $x_{\rm K}<0.6$.
This kind of precision will reveal the difference between the pion PDF
and kaon PDF in the valence and sea quark regions.
The gluon distribution of the kaon also can be extracted
with the scaling violation described by QCD evolution equations,
thanks to the wide $Q^2$ coverage of EicC.
The future leading $\Lambda$ tagged DIS experiment
will provide a lot of details of the interplay between EHM mechanism and HB mechanism,
by determining the strange valence quark distribution in the kaon,
which has a much larger coupling to the Higgs boson compared
to up quark or down quark.
In summary, the future EicC experiment on kaon structure has the impacts
to reveal the nature of the quasi Nambu-Goldstone particle in QCD,
to answer why the kaon mass is small (compared to hyperon),
and to test the nonperturbative predictions such as LQCD and DSE calculations.

\begin{acknowledgments}
We thank Prof. Craig D. Roberts for suggestions and discussions.
This work is supported by the Strategic Priority Research Program of Chinese Academy of Sciences under the Grant NO. XDB34030301,
the National Natural Science Foundation of China under the Grant NO. 12005266
and the Guangdong Major Project of Basic and Applied Basic Research under the Grant No. 2020B0301030008.
\end{acknowledgments}

\bibliographystyle{apsrev4-1}
\bibliography{refs}

\end{document}